\documentclass[journal]{IEEEtran}
\usepackage[utf8]{inputenc}
\usepackage{float}
\usepackage{xcolor}
\usepackage{color, soul}
\usepackage{enumerate}
\usepackage[bottom]{footmisc}
\usepackage{hyperref}
\usepackage{tabu}
\usepackage{caption}
\usepackage{bm, bbm}
\usepackage{mathtools}
\usepackage{amsmath, amssymb}
\usepackage{nicefrac}
\usepackage{empheq}

\usepackage{booktabs}
\usepackage{graphicx}
\usepackage{multirow}
\usepackage{units}
\usepackage{siunitx}
\usepackage{todonotes}
\usepackage{graphicx}
\usepackage{subcaption}
\usetikzlibrary{arrows.meta}
\usetikzlibrary{shapes,arrows}
\usetikzlibrary{automata, positioning}

\usepackage{amsmath,amsfonts,amsthm}

\title{A Markov Process Approach to Ensemble Control of Smart Buildings }
\author{Roman Pop $^{(a,b)}$, Ali Hassan $^{(b)}$, Kenneth Bruninx $^{(c)}$, Michael Chertkov $^{(1,d)}$ and Yury Dvorkin $^{(b)}$\\
$^{(a)}$ Skolkovo Institute of Science and Technology, Moscow, Russia\\
$^{(b)}$ Tandon School of Engineering,
NYU, New York, NY, USA \\
$^{(c)}$ KU Leuven Energy Institute, Leuven, Belgium \\
$^{(d)}$ T-4 and CNLS, Los Alamos National Laboratory, Los Alamos, NM, USA\vspace{-1cm}}

\date{\today}
\begin{document}

\maketitle

\begin{abstract}
This paper describes a step-by-step procedure that converts a physical model of a building into a Markov Process that characterizes energy consumption of this and other similar buildings. Relative to existing thermo-physics-based building models, the proposed procedure reduces model complexity and depends on fewer parameters, while also maintaining   accuracy and feasibility sufficient for system-level analyses. Furthermore, the proposed Markov Process approach makes it possible to leverage  real-time data streams available from intelligent data acquisition systems, which are  readily available in smart buildings,  and merge it with physics-based and statistical models. Construction of the Markov Process naturally leads to a Markov Decision Process formulation, which describes optimal probabilistic control of a collection of similar buildings. The approach is illustrated using validated building data from Belgium. 
\end{abstract}

\section{Introduction}

 Controllable electrical and heat appliances in residential and commercial buildings can provide significant flexibility to electrical and heat distribution systems by adjusting their consumption to help meeting operational system limits and alleviate overloads.  According to the US Federal Energy Regulatory Commission, the potential peak load reduction from all U.S. retail demand response programs was roughly 31,000 MW/year \cite{ferc_2017}. In the California Independent System Operator market, the total amount of proxy demand response capacity registered in 2016 was about 160 MW \cite{caiso_2016}. Currently, the ability of buildings to change their consumption is determined using field trips, remote on/off inspections, and equipment specifications \cite{7416149}. These practices  are labor-intensive and provide coarse, static, often weather-insensitive estimates. As a result, the true flexibility of buildings is often misestimated. 

Assessing flexibility that each building can provide requires accounting for electric power and heat dynamics, which are driven by comfort and behavioral preferences of occupants.  Currently, there are two large groups of methods to  model and forecast building electricity and heat consumption: (i) modeling relevant physical processes (e.g. heat transport, electro-mechanical considerations, Kirchoff's laws, evaporation, etc) and data-driven (e.g. statistical analyses and inference).  Physical models not only use available measurements and static building parameters, e.g. location, floor area, number of stories, detailed information on the heating, ventilation, and air conditioning (HVAC) system, lights, coils, doors and windows, but also operate with specific models that  govern dynamics of relevant characteristics \cite{COAKLEY2014123}. EnergyPlus, for example, is a popular simulation tool for modeling energy needs of buildings using detailed thermo- and mass- modeling of energy flows inside the building \cite{CRAWLEY2001319}. EnergyPlus can also be used for an off-line and off-site analyses to determine set point adjustments of the energy consumption \cite{osti_1247580}. The advantage of using the physics-based models is in their ability to describe buildings without prior observations. However, the performance of these models is highly sensitive to the number and accuracy of the underlying modeling choices ans assumptions, as well as to input parameters. Physics-based models often require more inputs than existing data acquisition systems can provide \cite{COAKLEY2014123}, and therefore incur significant uncertainties in both model parameters and dynamic processes. Using such models for controlling an ensemble of buildings may lead to computational issues that would prevent their scalability and implementation for real-life decision-making.  Due to these shortcomings, it is common to sacrifice modeling accuracy of  the physics-based models, which may lead to a loss of their predictive power.

On the other hand, in lieu of the physics-based models, one can use machine learning and statistical modeling to perform data-driven studies of buildings using a vast amount of historical data available at the buildings equipped with smart meters. These models are trained using the historical energy consumption data and other parameters (e.g. weather conditions, daily operational schedules, and control functionality) \cite{7416149, osti_1247580}. Then, the models can be used continuously to learn and predict energy usage from previously observed conditions. Availability of data is crucial for such approaches, especially when attempting to predict consumption with a minimum set of required inputs \cite{7416149}. Notably, such data is publicly available at an urban scale. For example, New York City’s Local Law 84 (LL84) requires that all commercial (including multi-family) buildings of 50,000 square feet or more must report energy and water consumption on an annual basis. Although this data is very coarse, it has been used in a combination with other building information (e.g. year built, floor area, property-use type, occupancy) to develop more accurate data-driven building models \cite{14a7b2f066b74e8ba2ee9c503c262b0a}. On the other hand, the data-driven models are data-intensive and building-specific and require large amounts of data for re-training or re-calibration, even  when minor changes are made to the buildings. This hinders  scalability of the data-driven models and their ability  to represent an ensemble of buildings with varying characteristics. 
Furthermore, numerous studies have revealed that data-driven models may yield discrepancies (up to 100\%) between the models outputs and the observed data \cite{14a7b2f066b74e8ba2ee9c503c262b0a, doi:10.1177/0143624407077393}. To reduce the gap between the prediction and the actual performance, researchers carried out calibration studies to tune the various inputs to match the observations \cite{doi:10.1080/19401493.2012.723750, YIN2016222, EISENHOWER2012292}. Nonetheless, calibration is still an over-specified and under-determined problem due to a relatively large number of inputs and a few measurable outputs \cite{7416149}. 

Alternatively,  machine learning and data-driven techniques can be leveraged to inform physics-based building models. In this manuscript, we construct a Markov Process (MP) to represent the energy consumption of building appliances to assess the  building flexibility  for various applications in electric power and heat distribution systems.  First, the physical building model is used to characterize the MP using the probability transition matrix. Based on this MP, we formulate the Markov Decision Process (MDP) that can in turn be used to optimally  control electric and heat appliances within buildings either by the local utility, or third-party aggregators, or building managers. The ultimate objective of this paper is to combine the scalability of the MDP with the accuracy of the physics-based models.

% added value? 
\vspace{-0.2cm}
\section{MDP for Building Modeling}

This section describes an approach to reducing multi-parametric and computationally demanding physical building models to a MP model. First, Section~\ref{sec:mp_and_mdp} reviews the MP approach to model an ensemble of homogeneous electrical or heat appliances and describes how this MP representation can be used for non-disruptive ensemble control. Second,  Section~\ref{sec:benchmark} describes a physically accurate, state-space model of a building that serves as a benchmark for the proposed approach. Finally, Section~\ref{sec:mp_building} details reduction of the physically accurate, state-space model of Section~\ref{sec:benchmark} to the MP described in Section~\ref{sec:mp_and_mdp}.

\vspace{-0.2cm}
\subsection{MP Representation for Ensemble Control} \label{sec:mp_and_mdp} 

As explored in our previous work \cite{Chertkov2017,  8442447, 8515118}, an ensemble of homogeneous electrical and heat appliances can be modeled as an MP. The MP modeling  is advantageous due to a sufficiently high accuracy and fidelity provided without rendering prohibitive computational complexity. This paper, leveraging the MP representation from \cite{Chertkov2017,  8442447, 8515118}, assumes that all electrical and heat appliances within the ensemble are geographically co-located, i.e. placed near one another, and the ensemble includes a sufficiently large  number of appliances. These two assumptions are hardly restrictive for modeling real-life buildings,  especially in urban areas with relatively high population densities, where normally hundreds of appliances are installed. Under these two assumptions,  \cite{Chertkov2017,  8442447, 8515118} represent the  ensemble of electrical or heat appliances as a discrete-time and discrete-space MP that can be extended to an MDP to control the aggregated state of the  ensemble.

MP is considered over a finite time horizon with discrete time periods  $0 \leq t\leq T$ that constitute set  $\mathcal{T}$. Accordingly, each ensemble is represented as a MP with a given number of discrete states, as  illustrated in Fig.~\ref{mdp:states}.  These states are obtained, given operating range and limits of each appliance, by discretizing the range of electric power or heat consumptions within the ensemble. The set of possible states is defined as $\mathcal{A}$ and $\alpha\neq\beta \in \mathcal{A}  $ are indices of the individual states.   Using the MP representation in Fig.~\ref{mdp:states},  the probability that the ensemble  is observed in state  $\alpha$ at time $t$ is described as $\rho_{t}^{\alpha}$.  The temporal evolution of the ensemble is accounted for by  $\mathcal{\overline{P}}_{t,b}^{\alpha\beta}$, representing transition probability from the state $\beta$ at time $t$ to the state $\alpha$ at time $t+1$. Note that $\mathcal{\overline{P}}_{t,b}^{\alpha\beta}$  stands for default evolution of the ensemble, i.e. it is driven by consumption preferences and choices of consumers that use electrical and heat appliances, and does not include the effect of control actions of the utility or third-party aggregators.

Similarly to \cite{Chertkov2017,  8442447, 8515118}, the MP approach introduced above leads to the following MDP formulation describing control of an aggregated state of the ensemble:

\begin{figure}
\centering
  \begin{tikzpicture}[font=\footnotesize]
  % Add the states
  \node[state,fill=gray!20!white,minimum size=0.5cm] (s1) {1};
  \node[state,fill=gray!20!white,
     left=1cm of s1,minimum size=0.5cm] (s2) {2};
  \node[state,fill=gray!20!white,
     left=1cm of s2,minimum size=0.5cm] (s3) {3};
  \node[state,fill=gray!20!white,
     left=1cm of s3,minimum size=0.5cm] (s4) {4};
   \node[state,fill=gray!20!white,
     below=2cm of s4,minimum size=0.5cm] (s5) {5};
  \node[state,fill=gray!20!white,
     right=1cm of s5,minimum size=0.5cm] (s6) {6};
  \node[state,fill=gray!20!white,
     right=1cm of s6,minimum size=0.5cm] (s7) {7};
  \node[state,fill=gray!20!white,
     right=1cm of s7,minimum size=0.5cm] (s8) {8};
  \node[right=0.2cm of s1,font=\small,text width=1.5cm] {ON states};
  \node[right=0.2cm of s8,font=\small,text width=1.5cm] {OFF states};
  \coordinate[below of=s5, yshift=0.15cm] (c1);
  \coordinate[below of=s8, yshift=0.15cm] (c2);
  % Connect the states with arrows
  \draw[>=latex,every loop,fill=black!70,
    draw=black!70,
     auto=right,
     line width=0.5mm]
    (s1) edge[line width=1mm] (s2)
    (s2) edge[line width=1mm] (s3)
    (s3) edge[line width=1mm] (s4)
    (s4) edge[line width=1mm] (s5)
    (s5) edge[line width=1mm] (s6)
    (s6) edge[line width=1mm] (s7)
    (s7) edge[line width=1mm] (s8)
    (s8) edge[line width=1mm] (s1)
    (s1) edge[loop above,line width=0.7mm]  (s1)
    (s1) edge[bend right, auto=right] (s3)
    (s1) edge[bend right, auto=right,looseness=1.3,line width=0.1mm] (s4)
    (s1) edge[line width=0.1mm] (s5)
    (s1) edge[line width=0.3mm] (s6)
    (s1) edge[line width=0.5mm] (s7)
    (s1) edge[bend left, auto=left,line width=0.1mm] (s8);
  \draw [-,thick,dashed, line width=0.25mm] (s5) -- (c1) node [near start] {};
  \draw [-,thick,dashed, line width=0.25mm] (s8) -- (c2) node [near start] {};
  \draw [{<[scale=1.1]}-{>[scale=1.1]},thick, line width=0.25mm,anchor=south] (c1) -- (c2) node [midway] {Dispatch range} node [very near start,anchor=north east] {} node [very near end,anchor=north west] {};
  
  \end{tikzpicture}
  \vspace{4pt}
  \caption{A Markovian representation of the  ensemble of electrical or heat appliances with  eight discrete states displaying all possible transitions from state $1$, 
  \cite{8515118}. }
  \label{mdp:states}
\end{figure}
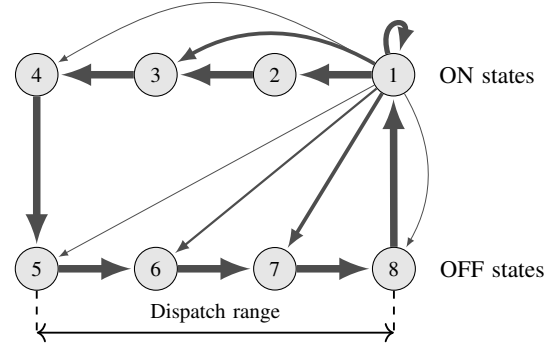

% Parapgraph about MDP representation and discretization
% General notations. 

\allowdisplaybreaks
\begin{align}
&\underset{\substack{\rho,\mathcal{P}, p, q}}{\text{min}}\sum_{t \in \mathcal{T}}\mathbb{E}_{\rho}
\sum_{t \in \mathcal{T}} \! \sum_{\alpha \in \mathcal{A}} \! \left(-U_{t+1}^{\alpha} + \! \textcolor{black}{\sum_{\beta \in \mathcal{A}}} \gamma_{t}^{\alpha\beta} \log\! \frac{\mathcal{P}_{t}^{\alpha\beta}}{\overline{\mathcal{P}}_{t}^{\alpha\beta}}\right) 
%\rho_{t,b}^{\beta} 
\label{MDP:obj}
\end{align}
subject to:
\begin{align}
&\rho_{t+1}^{\alpha} = \sum_{\beta \in \mathcal{A}} \mathcal{P}_{t}^{\alpha\beta} \rho_{t}^{\beta},  \quad \forall \alpha \in \mathcal{A}, t \in \mathcal{T} %\backslash |N_\mathcal{T}|
\label{MDP_evol} \\
& p_{t}= \sum_{\alpha \in \mathcal{A}} p^{\alpha} \rho_{t}^{\alpha}, \quad \forall t \in \mathcal{T},  \label{mdp_injP1} \\
& q_{t} = \sum_{\alpha \in \mathcal{A}} q^{\alpha} \rho_{t}^{\alpha} ,  \quad \forall t \in \mathcal{T}, \label{mdp_injQ1} \\
&\sum_{\alpha \in \mathcal{A}} \mathcal{P}_{t}^{\alpha \beta} = 1,  \quad \forall t \in \mathcal{T}, \beta \in \mathcal{A}, \label{mdp_integrality} 
%&\sum_{\alpha} \mathcal{P}_{i,t}^{\alpha\beta} = 1
\end{align}
\allowdisplaybreaks[0]
where $\mathcal{P}_{t}^{\alpha\beta}$, $\rho_{t}^{\alpha}$, $\rho_{t}^{\beta}$, $p_{t}$ and $q_{t}$  are decision variables,  $\overline{\mathcal{P}}_{t}^{\alpha\beta}$,  $U_{t+1}^{\alpha}$, $p_{t}^\alpha$ and $q_{t}^\beta$ are given parameters, and operator $\mathbb{E}_{\rho}$ stands for evaluating expectations over $\rho$. Eq.~\eqref{MDP:obj}-\eqref{mdp_integrality} describes optimal evolution of the ensemble under control from state $\beta$ at time $t$ to state $\alpha$ at time $t+1$. Eq.~\eqref{MDP:obj} is the objective function that includes two terms. The first term maximizes the total utility of the ensemble observed in state $\alpha$  at time $t+1$, denoted as $U_{t+1}^{\alpha}$, while the second term introduces cost of the discomfort due to control decisions ${\mathcal{P}}_{t}^{\alpha\beta}$. Decisions ${\mathcal{P}}_{t}^{\alpha\beta}$ represent the state transitions that are requested by the utility or third-party aggregator and, therefore, are likely to deviate from the default evolution  described by $\overline{\mathcal{P}}_{t}^{\alpha\beta}$. Naturally, the difference in transitions between the requested and default states, i.e. between  ${\mathcal{P}}_{t}^{\alpha\beta}$ and $\overline{\mathcal{P}}_{t}^{\alpha\beta}$, measures the cost of discomfort computed using the  Kullback-Leibler (KL) divergence\footnote{The choice of the Kullback-Leibler divergence for the discomfort cost is motivated by its wide use for modeling randomness of discrete and continuous time series. Other penalty functions can also be used instead.} in the second term of the objective function using penalty $\gamma^{\alpha\beta}$. Respectively, if ${\mathcal{P}}_{t}^{\alpha\beta}=\overline{\mathcal{P}}_{t}^{\alpha\beta}$, there is no discomfort cost and the second term in Eq.~\eqref{MDP:obj} is equal to zero. Eq.~\eqref{MDP_evol} describes temporal evolution of the ensemble from $t$ to $t+1$, based on the states and transitions illustrated in Fig.~\ref{mdp:states}. Eqs.~\eqref{mdp_injP1}-\eqref{mdp_injQ1} compute the expected active and reactive power injections of the TCL ensemble into the distribution system, where parameters $p^{\alpha}$  and $q^{\alpha}$ describe active and reactive power consumptions in the  state $\alpha$. Eq.~\eqref{mdp_integrality} imposes the integrality constraint on the control  decisions  ${\mathcal{P}}_{t}^{\alpha\beta}$ such that their total probability is equal to one (normalized). 

For the sake of simplicity, in the following we set $\gamma^{\alpha\beta}_t=1, \forall \alpha\neq\beta \in \mathcal{A}, t \in \mathcal{T}$, which converts the MDP in Eq.~\eqref{MDP:obj}-\eqref{mdp_integrality} to a Linearly Solvable MDP that can be solved analytically (see \cite{Todorov2006}).

% Model description
\subsection{Physical Building Model} \label{sec:benchmark}

Default transition probabilities $\overline{\mathcal{P}}_{t}^{\alpha\beta}$ can be viewed as a reduced-form representation of the actual changes in electrical and heat measurements observed in a real-life building or as an aggregated consumption metric of a detailed, physical building model. Therefore, if one represents a physical model of a building as $\overline{\mathcal{P}}_{t}^{\alpha\beta}$, the MDP described by Eq.~\eqref{MDP:obj}-\eqref{mdp_integrality} can be used to control the ensemble of appliances within a given building.   This section explains the physical building model used below to construct  $\overline{\mathcal{P}}_{t}^{\alpha\beta}$ for a given building.

The physical building model used in this paper is a dynamic, linear, state-space model of a building and its heating system constructed based on \cite{Reynders2014}. This model captures  thermal dynamics of a given building  envelope and its heating system using a nine-state representation over the two thermal zones (e.g. differentiating between day and night). Accuracy of the model is validated via  detailed physical simulations  using the IDEAS library \cite{DeConinck2014} in Modelica, as described in \cite{Patteeuw2014}. Furthermore, as shown in \cite{Reynders2014} and \cite{Patteeuw2014}, this linear state-space model approximates  nonlinear dynamics of residential buildings with heat pumps with a sufficient level of accuracy for system-level studies, i.e. from the perspective of electrical or heat distribution systems.  For example, these models have been used to access controllability of an ensemble of thermostatically controlled loads in \cite{bruninx2018valuing}.

The building heating system consists of an air-coupled heat pump (HP) and a back-up electric resistance heater (A), which supply heat to the floor heating system in the day and night zones, i.e. space heating (SH), and to the storage tank for domestic hot water (HW) \cite{Patteeuw2014,Arteconi2015}. Under the assumption of perfectly known external and internal states and gains (e.g. outside temperature, solar radiance, temperature inside the building) and user-defined settings (i.e., temperature bounds $\underline{T_{pj}^{SH}}$ and $\overline{T_{pj}^{SH}}$), the  consumption of the heat pump  ($P^{HP}_{t}$) and auxiliary heater ($P^{A}_{t}$)  for space heating and hot water production can be minimized for  each building $j \in \mathcal{J}$, where $\mathcal{J}$ is a set of buildings, using the optimization problem formulated in  \cite{bruninx2018valuing} and stated below:
\begin{equation}
    \min  \sum_{t\in\mathcal{T}} d^{\mathrm{H}}_{j,t}  \label{el_power}
\end{equation}
subject to 
\allowdisplaybreaks
\begin{align}
& \forall t \in \mathcal{T}: \;\;\;   d^{\mathrm{H}}_{j,t} = p^{\mathrm{HP}}_{j,t} + p^{\mathrm{A}}_{j,t} \label{eq:DR_1} \\
& \forall t \in \mathcal{T}: \;\;\;    p^{\mathrm{HP}}_{j,t} = p^{\mathrm{HP,SH}}_{j,t} + p^{\mathrm{HP,HW}}_{j,t} \leq \overline{P^{\mathrm{HP}}_{j}}  \label{eq:DR_2} \\
& \forall t \in \mathcal{T}: \;\;\;   p^{\mathrm{A}}_{j,t} = p^{\mathrm{A,SH}}_{j,t} + p^{\mathrm{A,HW}}_{j,t} \leq \overline{P^{\mathrm{A}}_{j}} \label{eq:DR_3} \\
& \forall t \in \mathcal{T}: \;\;\;  \sum_{z \in \mathcal{Z}} \dot{q}^{\mathrm{SH}}_{j,t,z} = COP^{\mathrm{SH}}_{h} \cdot p^{\mathrm{HP,SH}}_{hj} + p^{\mathrm{A,SH}}_{hj} \label{eq:DR_4} \\
&\forall p \in \mathcal{P}, \forall t \in \mathcal{T}: \;\;\;  t^{\mathrm{SH}}_{j,p,t}  = A_{j,p} \cdot t^{\mathrm{SH}}_{j,p,t-1} \label{eq:state-space} \\
&\hspace{0.75cm} + \sum_{z \in \mathcal{Z}} B_{j,p,z} \cdot \dot{q}^{\mathrm{SH}}_{\mathrm{j,t,z}}  +  E^{\mathrm{SH}}_{j,p,t}  \nonumber \\
& \forall p \in \mathcal{P}, \forall t \in \mathcal{T}: \;\;\;  \underline{T^{\mathrm{SH}}_{j,p,t}} \leq t^{\mathrm{SH}}_{j,p,t}  \leq \overline{T^{\mathrm{SH}}_{j,p,t}} \label{eq:comfort_constraint} \\
& \forall t \in \mathcal{T}: \;\;\;   t^{\mathrm{HW}}_{j,t} =  t^{\mathrm{HW}}_{j,t-1} - \frac{TP \cdot G_j}{C_j} \cdot (t^{\mathrm{HW}}_{j,t}-T^{\mathrm{E}}) \label{eq:hot_water_storage} \\
& \hspace{0.75cm} + \frac{TP}{C_j} \cdot \left(COP^{\mathrm{HW}}_{j} \cdot p^{\mathrm{HP,HW}}_{j,t} + p^{\mathrm{A,HW}}_{j,t} - \dot{Q}^{\mathrm{D}}_{j,t} \right) \nonumber  \\
& \forall t \in \mathcal{T}: \;\;\;  \underline{T^{\mathrm{HW}}_{j,t}} \leq t^{\mathrm{HW}}_{j,t} \leq \overline{T^{\mathrm{HP}}_{j,t}} \label{eq:hot_water_constraint}
\end{align}
\allowdisplaybreaks[0]
Eq.~\eqref{eq:DR_1} aggregates the electrical power required by the heat pump ($p^{\mathrm{HP}}_{j,t} $) and auxiliary heater ($p^{\mathrm{A}}_{j,t}$), which are limited to their nameplate capacity $\overline{P^{\mathrm{HP}}_{j}}$ and $\overline{P^{\mathrm{A}}_{j}}$ in \eqref{eq:DR_2} and \eqref{eq:DR_3}. We distinguish between the electrical power required for space heating ($p^{\mathrm{HP,SH}}_\mathrm{{j,t}} \geq 0$ and $p^{\mathrm{A,SH}}_{j,t} \geq 0$)  and hot water production ($p^{\mathrm{HP,HW}}_{j,t} \geq 0$ and $p^{\mathrm{A,HW}}_{j,t} \geq 0$) supplied by the heat pumps and auxiliary heaters. The electrical power consumed by the heat pump and the auxiliary heater is related to the required thermal power for space heating $\dot{q}^{\mathrm{SH}}_{j,t,z} \geq 0$ in each zone $z$ (set $\mathcal{Z}$) using the coefficient of performance $COP^{\mathrm{SH}}_{j}$ of each building $j$ as given by Eq. ~\eqref{eq:DR_4}.  Eq.~\eqref{eq:state-space} describes the temperature evolution ($t^{\mathrm{SH}}_{j,p,t} \geq 0$) in each temperature state $p$ (set $P$) based on the  thermal power supplied to each building zone ($\dot{q}^{\mathrm{SH}}_{j,t,z}$),  thermal losses, internal and external gains $E^{\mathrm{SH}}_{j,p,t}$. Equation \eqref{eq:comfort_constraint} enforces the user-defined comfort constraints on the indoor air temperature ($\underline{T^{\mathrm{SH}}_{j,p,t}}$, $\overline{T^{\mathrm{SH}}_{j,p,t}}$). 
Eq.~\eqref{eq:hot_water_storage} describes evolution of temperature $t^{\mathrm{HW}}_{j,t}$ in the hot water storage tank, which depends on the thermal conductance of the storage tank, $G_j$, the temperature of its surroundings $T^{\mathrm{E}}$, the thermal power supplied to the hot water storage tank, the thermal capacity of the storage tank and its contents $C_j$. $\dot{Q}^{\mathrm{D}}_{j,t}$ represents the withdrawal of thermal power from the hot water storage tank due to the hot water consumption, which follows a pre-defined, consumer-specific profile. $TP$ is the temporal resolution of the model. Equation \eqref{eq:hot_water_constraint} ensures the user-defined availability of hot water at temperatures between $\underline{T^{\mathrm{HW}}_{j,t}}$ and $\overline{T^{\mathrm{HP}}_{j,t}}$. 

\subsection{From building data to MP} \label{sec:mp_building}

\begin{figure}[!htpb]
    \centering
    \includegraphics[width=\linewidth]{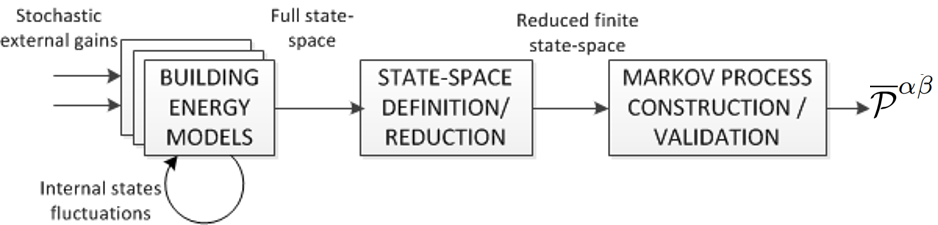}
    \caption{The proposed three-step procedure to construct MP for optimal ensemble control of buildings.}
    \label{flowchart}
\end{figure}

Given the physically accurate building model discussed in Section~\ref{sec:benchmark}, the goal of this section  is to describe reduction of  the optimization in Eq.~\eqref{el_power}-\eqref{eq:hot_water_constraint} to MDP in Eq.~\eqref{MDP:obj}-\eqref{mdp_integrality}.  The reduction is illustrated in Fig.~\ref{flowchart}. It  includes the following three steps: (i)  building data generation and aggregation, (ii) state-space definition and reduction, and (iii) construction and validation of the resulting MP. These steps are further detailed below.

\subsubsection{Step 1: Building Energy Models}  This step applies the physical building model in Eq.~\eqref{el_power}-\eqref{eq:hot_water_constraint} to a given set of buildings to assess their performance under different use scenarios and  external and internal gains.  To simplify the physical modeling process, we consider an ensemble of physically homogeneous buildings with different occupancy levels and comfort constraints.  Each  building in the ensemble is influenced by stochastic, external gains, representing outside temperature, solar radiance, wind etc. Similar in their physical properties and exposed to similar climatic geographical conditions, buildings may be described by a slightly different set of parameters and they may also be subject to some external (e.g. weather-driven) and internal (e.g. windows opened/closed, thermostat set-point changes due to user preferences, fluctuations in occupancy etc.) fluctuations. To account for the internal fluctuations, the physical building model was simulated for different occupancy profiles, provided via binary sequence of ``presence/absence'' states of occupants. These simulations provide  a state-space representation, which includes stochasticity in the internal gains.

\subsubsection{Step 2: State-space definition and reduction}

Given a full set of data obtained using the simulations described in Step 1, as well as available building measurements over a given  finite time horizon, one constructs a high-dimensional state-space model to control an ensemble of buildings. However,  in practice it is not feasible to monitor and capture all state parameters and, therefore, we derive a $d$-dimensional state-space $S \in \mathbbmtt{R}$, where $d$ is a user-defined number which is smaller than the degree of the original building model. Note that the degree of this reduction depends on the original physical model (state variables and their number) and also on the tolerance to modeling inaccuracies. Therefore, in every case, this decision should be made cautiously based on the intended model application. In the following, we will use a  two-state model ($d=2$) to simplify visualization of our results. From the physical perspective, the reduced states should constitute  a meaningful physical process. Thus, changing a one-state variable should  result in changes in another state variable. Therefore, given the optimization described in Eq.~\eqref{el_power}-\eqref{eq:hot_water_constraint}, it is reasonable to select the inside temperature ($T_{in}$) and the electric power for heating ($P^{HP}_{j,t}$ or $P^A_{j,t}$) because these state variables are mutually dependent (i.e. an increase in electric power should result in a temperature increase). Furthermore, these variables are of importance to consumers since $T_{in}$ defines comfort preferences and $P^{HP}_{j,t}$ and $P^A_{j,t}$ affect the electricity bill.

Given these selected state variables, we construct the reduced state-space building model by uniformly discretizing each state variable into numerically ordered ranges to represent their dispatch range, e.g. as shown in Fig.~\ref{mdp:states}. For example, the range for $T_{in}$ is represented by  $n$ states  and the range for $P^{HP}_{j,t}$ or $P^A_{j,t}$ is represented by $m$ states. Then, the list of possible states will be represented by all possible cross-combinations of these states for $T_{in}$, $P^{HP}_{j,t}$ or $P^A_{j,t}$, i.e.  $n \times m$. Under these assumptions, $\overline{\mathcal{P}}_{t}^{\alpha\beta}$  and ${\mathcal{P}}_{t}^{\alpha\beta}$ will also have dimensions $n \times m$.

\subsubsection{Step 3: Construct the MP}

Finally, we construct the transition probability matrix $\overline{\mathcal{P}}_{t}^{\alpha\beta}$, which describes the steady-state evolution of the system. Given a reduced finite state-space representation obtained in Step 2, the transition probabilities $\overline{P}_{\alpha\beta}$ are computed for  all states by tracing the number of transitions between states $\beta$ and $\alpha$ for all available simulations regardless of their duration and time interval. The resulting $d$-dimensional array should then be normalized to meet the integrality criteria of the MP  setting. 

Note that this approach for constructing the MP may lead to some non-physical states, which cannot be implemented in practice. To avoid such infeasibilities, respective elements of the matrix $\overline{P}_{\alpha\beta}$ are set to 0.

\section{Case study}

This case study aims to illustrate how the proposed MP approach represents an ensemble of buildings and to demonstrate its utility for non-intrusive control of appliances in the building.

\subsection{Data for Constructing Physical Building Models}
We construct a physical building model,  described in Section~\ref{sec:benchmark}, using  parameters for individual buildings, their heating systems, occupancy and behavioral choices of occupants reported in \cite{Reynders2014,Protopapadaki2014,Baetens2015,Patteeuw2014,Patteeuw2014JP}. The key parameters are also summarized below. 

We consider an \textquoteleft average' low-energy building, in which the day  and night zones have a surface area of $132$ $m^2$ and $138$ $m^2$, respectively,   \cite{cyx2011iee}. All these buildings are assumed to have undergone a renovation of windows, air tightness, walls, floor and roof resulting in  an average U-value of $0.3$ $W/K$ and a ventilation rate of $0.4$ $ACH$ (air changes per hour). The heat pump is sized to meet $80 \%$ of the peak heat demand and the remainder is assumed to be supplied by the back-up electric resistance heater. As in \cite{EcofysStudyAppendix2013,Verhelst2012}, the COP of the heat pump is calculated assuming a nominal supply water temperature of the floor heating of $35\,^{\circ}C$. The hot water storage tanks have the volume of either $200$ $l$ or $300$ $l$ (vary based on the maximum daily hot water demand at $50\,^{\circ}C$). The maximum supply temperature of the heat pump is $60\,^{\circ}C$. 

When occupants are in the building, the lower bounds for the indoor temperature set points are $20 \,^{\circ}C$ and $18 \,^{\circ}C$ for the day  and night zones, respectively, while the upper bounds are $22 \,^{\circ}C$ and $20 \,^{\circ}C$, respectively \cite{Peeters2009}. When unoccupied, the lower bound is set to $16 \,^{\circ}C$ and the upper bound is unchanged. To account for the stochasticity of internal gains, we generate 52 user behavior profiles using  STroBE (\url{https://github.com/open-ideas/StROBe}), a statistical tool described in \cite{Baetens2015}, which allows to capture different levels of consumers' presence (e.g. occupied or not). External gains are modeled using weather data for Uccle,  Belgium metered in 2013. The weather measurements include solar gains computed using the model in \cite{Baetens2012}. The heating load was simulated during four weeks of calendar year 2013 with a 15 minute-time step.

\subsection{Construction of MP} \label{sec:results_mp}

As explained above, we construct the MP using two state-space variables. In this representation,  the power variables, either electric power of the heat pump ($p^{HP}_{j,t}$) or auxiliary heater ($p^{A}_{j,t}$), are considered as independent variables, while the inside building temperature ($T_{in}^p$ ) is considered as a dependent variable.  This selection of dependent and independent variables allows to use the model of Section~\ref{sec:benchmark} that establishes a connection between the temperature and power variables. 

Given this modeling assumptions we have experimented with the following cases:

\begin{itemize}
    \item[] \textbf{Case 1}: Construct MP for the power consumption of the heat pump ($p^{HP}_{j,t}$), while ignoring the auxiliary heater ($p^{A}_{j,t}$).
    \item[] \textbf{Case 2}: Construct MP for the power consumption of the auxiliary heater ($p^{A}_{j,t}$), while ignoring the heat pump ($p^{HP}_{j,t}$).
\end{itemize}

This split in two cases is motivated by the different operating principles and low-level controls available for the heat pumps and auxiliary heaters that can affect their operations and, thus, thermal dynamics in the buildings. Note that in both cases, the inside building temperature ($T_{in}^p$) is a dependent variable.

% figures -> 
\begin{figure}[!b]
    \centering
    \includegraphics[width=0.7\linewidth]{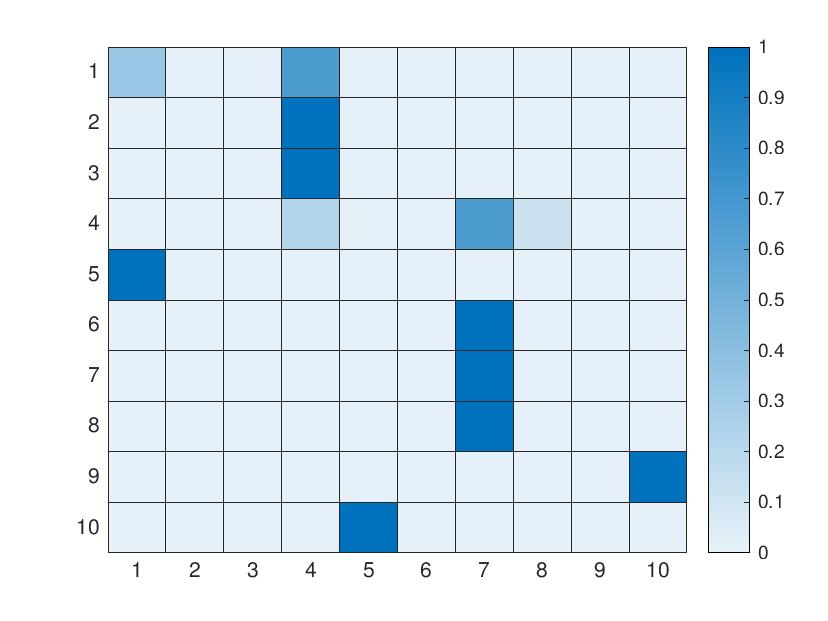}
    \caption{Transition probability matrix $\overline{\mathcal{P}}^{\alpha\beta}$ with  10 states, i.e. $\text{card}(\mathcal{P})=10$, where $p^{HP}_{j,t}$ is modeled as an independent variable. The color density indicates the probability value as shown in the vertical sidebar. }
    \label{case1_2}
\end{figure}

\begin{figure}[!b]
    \centering
    \includegraphics[width=0.7\linewidth]{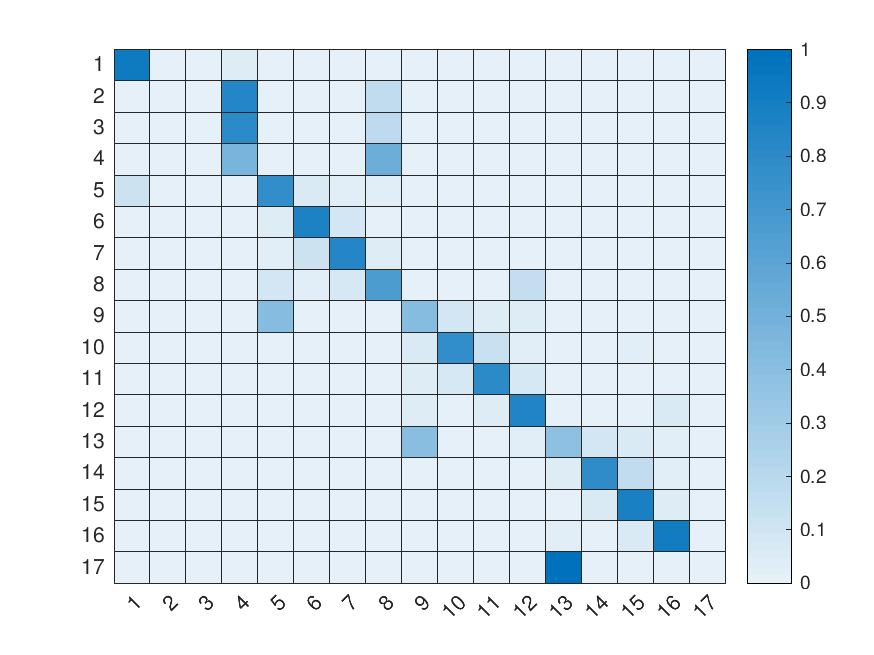}
    \caption{Transition probability matrix $\overline{\mathcal{P}}^{\alpha\beta}$ with  17 states, i.e. $\text{card}(\mathcal{P})=17$, where $p^{HP}_{j,t}$ is modeled as an independent variable. The color density indicates the probability value as shown in the vertical sidebar.}
    \label{case1_1}
\end{figure} 

\subsubsection{Case 1: MP for $p^{HP}_{j,t}$} \label{sec:results_case1}

% The resulting MP matrix is constructed of 5 and 8 states for $T_{indoor}$ and 2 states for occupancy - total 10 or 16 states. It is also averaged through the 52 different occupancy profiles.

Based on our experiments, we established that the most physically accurate parameters of the MP are obtained for the cases when the range of feasible values on the inside temperature is uniformly discretized in either 10 or 17 states, i.e. $\text{card}(\mathcal{P})=10$ or $\text{card}(\mathcal{P}) =17$. In this case, the range of  possible states for independent variable $p^{HP}_{j,t}$ is  also uniformly discretized in 10 or 17 states. Accordingly, the state space can be represented by $10\times10$ and $17\times17$ states, respectively, and  matrix $\overline{\mathcal{P}}^{\alpha\beta}$ has dimensions of $10\times10$ and $17\times17$, respectively.

Using the physical model and occupancy data of buildings described above, we obtain transition probability matrices  $\overline{\mathcal{P}}^{\alpha\beta}$  with dimensions $10\times10$ and $17\times17$. These transition probability matrices are shown in Fig.~\ref{case1_2} and Fig.~\ref{case1_1}, respectively. Note that as the number of states increases, the relative weight of elements on the main diagonal increases, while the relative weight of the off-diagonal elements reduces. Also note that increasing the number of states leads to a sparser matrix. This sparsity indicates a fairly small number and magnitude of exogenous disturbances under normal building operations. If such disturbances are more frequent and larger due to internal or external gains (e.g. natural disasters), the transition matrix would become less sparse. 

\begin{figure}[!t]
    \centering
    \includegraphics[width=0.7\linewidth]{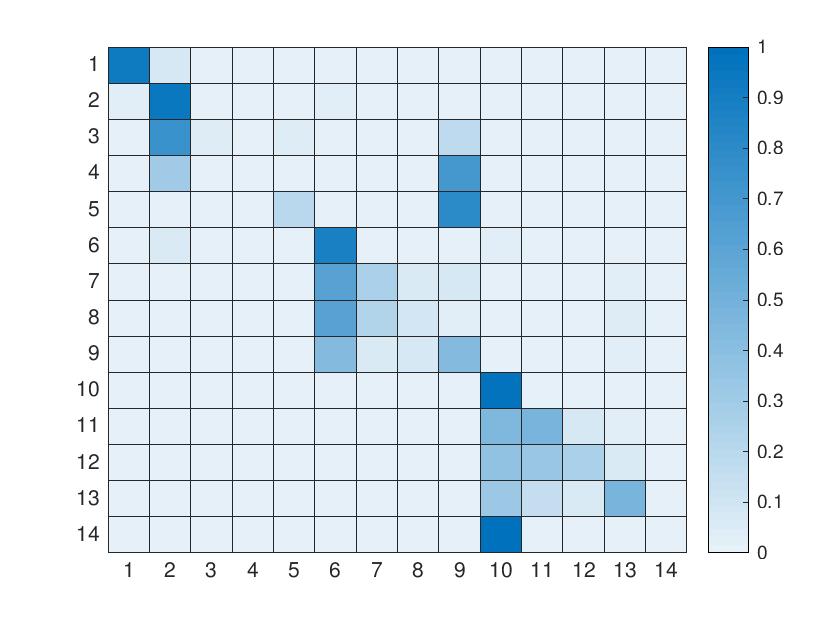}
    \caption{Transition probability matrix $\overline{\mathcal{P}}^{\alpha\beta}$ with  14 states, i.e. $\text{card}(\mathcal{P})=14$, where $p^{A}_{j,t}$ is modeled as an independent variable. The color density indicates the probability value as shown in the vertical sidebar.}
    \label{case2_1}
\end{figure}

\begin{figure}[!t]
    \centering
    \includegraphics[width=0.7\linewidth]{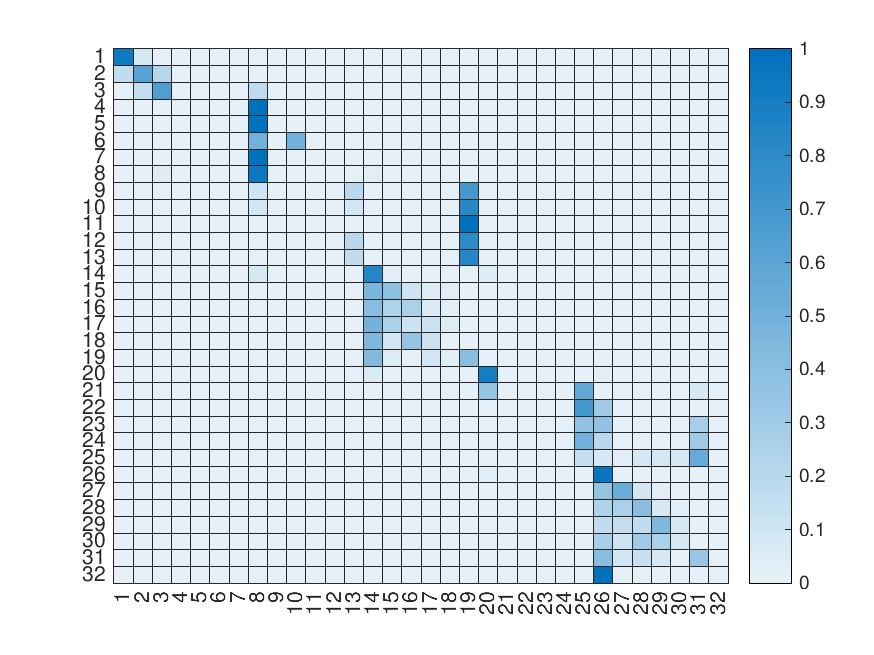}
    \caption{Transition probability matrix $\overline{\mathcal{P}}^{\alpha\beta}$ with  32 states, i.e. $\text{card}(\mathcal{P})=32$, where $p^{A}_{j,t}$ is modeled as an independent variable. The color density indicates the probability value as shown in the vertical sidebar.}
    \label{case2_2}
\end{figure}

\subsubsection{Case 2: MP for $p^{A}_{j,t}$} We repeat the process of Section~\ref{sec:results_case1} and obtain the MP for the case with  $p^{A}_{j,t}$ modeled as an independent variable. In this case, our experiments suggest that the most physically accurate parameters of the MP are obtained for the cases when the range of feasible states is discretized in 14 and 32 states, respectively. The resulting transition probability matrices are illustrated in Fig.~\ref{case2_1} and Fig.~\ref{case2_2}. Comparing to results in Fig.~\ref{case1_2} and Fig.~\ref{case1_1}, we note that the individual elements of the matrices obtained with different independent variables vary significantly.

\subsection{MDP for Ensemble Control}

\begin{figure}[!t]
    \centering
    \includegraphics[width=\linewidth]{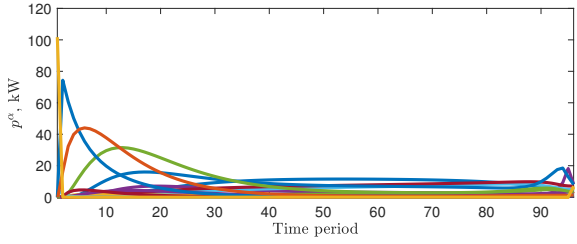}
    \caption{Optimized power consumption $p^{\alpha}$ obtained by solving the MDP in Eq.~\eqref{MDP:obj}-\eqref{mdp_integrality} with $\overline{\mathcal{P}}^{\alpha\beta}$ from   Figure~\ref{case1_1}. Different colors denote different states $\alpha \in \mathcal{A}$. }
    
    \label{mdp_case1_1}
\end{figure}

\begin{figure}[!t]
    \centering
    \includegraphics[width=\linewidth]{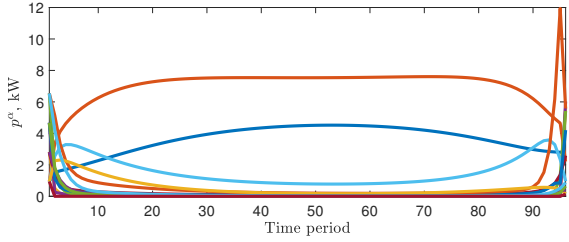}
    \caption{Optimized power consumption $p^{\alpha}$ obtained by solving the MDP in Eq.~\eqref{MDP:obj}-\eqref{mdp_integrality} with $\overline{\mathcal{P}}^{\alpha\beta}$ from   Figure~\ref{case2_1}. Different colors denote different states $\alpha \in \mathcal{A}$.}
    
        \label{mdp_case2_1}
\end{figure}

The MP constructed in Section~\ref{sec:results_mp} can now be used to implement the control of building ensembles as described in Section~\ref{sec:mp_and_mdp}. Using the solution procedure described in \cite{8515118}, we solve the optimization problem in Eq.~\eqref{MDP:obj}-\eqref{mdp_integrality}  with the values of $\overline{\mathcal{P}}^{\alpha\beta}$ described in Fig.~\ref{case1_1} and Fig.~\ref{case2_1} and  the optimization horizon  set to 96 hourly time intervals. Optimal decisions ($\mathcal{P}_{t}^{\alpha\beta}$) for each case lead to the active power consumption  displayed in Fig.~\ref{mdp_case1_1} and Fig.~\ref{mdp_case2_1}, respectively. Similarly to the observations \cite{8515118}, these optimal decisions change over time following  changing prices  during the optimization horizon. Note that the results in Fig.~\ref{mdp_case1_1} and Fig.~\ref{mdp_case2_1} demonstrate that the optimal decisions converge to one or may be a very few states due to more revenues opportunities in these states.

% \section{Conclusion}
% \textcolor{blue}{to be completed..}

\section*{Acknowledgements} The work at New York University was in part supported by NSF Award \# CMMI-1825212. The work at LANL was carried out under the auspices of the National Nuclear Security Administration of the U.S. Department of Energy under Contract No. DE-AC52-06NA25396. The work was partially supported by DOE/OE/GMLC and LANL/LDRD/CNLS projects.

\bibliographystyle{IEEEtran}
\bibliography{main}

\begin{thebibliography}{10}
\providecommand{\url}[1]{#1}
\csname url@samestyle\endcsname
\providecommand{\newblock}{\relax}
\providecommand{\bibinfo}[2]{#2}
\providecommand{\BIBentrySTDinterwordspacing}{\spaceskip=0pt\relax}
\providecommand{\BIBentryALTinterwordstretchfactor}{4}
\providecommand{\BIBentryALTinterwordspacing}{\spaceskip=\fontdimen2\font plus
\BIBentryALTinterwordstretchfactor\fontdimen3\font minus
  \fontdimen4\font\relax}
\providecommand{\BIBforeignlanguage}[2]{{%
\expandafter\ifx\csname l@#1\endcsname\relax
\typeout{** WARNING: IEEEtran.bst: No hyphenation pattern has been}%
\typeout{** loaded for the language `#1'. Using the pattern for}%
\typeout{** the default language instead.}%
\else
\language=\csname l@#1\endcsname
\fi
#2}}
\providecommand{\BIBdecl}{\relax}
\BIBdecl

\bibitem{ferc_2017}
\BIBentryALTinterwordspacing
{Federal Energy Regulatory Commission}, ``Assessment of demand response and
  advanced metering,'' 2017. [Online]. Available:
  \url{https://www.ferc.gov/legal/staff-reports/2017/DR-AM-Report2017.pdf}
\BIBentrySTDinterwordspacing

\bibitem{caiso_2016}
\BIBentryALTinterwordspacing
{California Independent System Operator}, ``2016 annual report on market issues
  \& performance,'' 2016. [Online]. Available:
  \url{http://www.caiso.com/Documents/2016AnnualReportonMarketIssuesandPerformance.pdf}
\BIBentrySTDinterwordspacing

\bibitem{7416149}
{T. Samad et al.}, ``Automated demand response for smart buildings and
  microgrids: The state of the practice and research challenges,''
  \emph{Proceedings of the IEEE}, vol. 104, no.~4, pp. 726--744, April 2016.

\bibitem{COAKLEY2014123}
D.~Coakley, P.~Raftery, and M.~Keane, ``A review of methods to match building
  energy simulation models to measured data,'' \emph{Renewable and Sustainable
  Energy Reviews}, vol.~37, pp. 123 -- 141, 2014.

\bibitem{CRAWLEY2001319}
{D. B. Crawley et al.}, ``Energyplus: creating a new-generation building energy
  simulation program,'' \emph{Energy and Buildings}, vol.~33, no.~4, pp. 319 --
  331, 2001, special Issue: BUILDING SIMULATION'99.

\bibitem{osti_1247580}
{D. Chassin et al.}, ``A new thermostat for real-time price demand response:
  Cost, comfort and energy impacts of discrete-time control without deadband,''
  \emph{Applied Energy}, vol. 155, 10 2015.

\bibitem{14a7b2f066b74e8ba2ee9c503c262b0a}
C.~Kontokosta, \emph{\BIBforeignlanguage{English (US)}{Predicting Building
  Energy Efficiency Using New York City Benchmarking Data}}, 2012.

\bibitem{doi:10.1177/0143624407077393}
F.~Karlsson, P.~Rohdin, and M.-L. Persson, ``Measured and predicted energy
  demand of a low energy building: important aspects when using building energy
  simulation,'' \emph{Building Services Engineering Research and Technology},
  vol.~28, no.~3, pp. 223--235, 2007.

\bibitem{doi:10.1080/19401493.2012.723750}
{A.T. Booth et al.}, ``A hierarchical bayesian framework for calibrating
  micro-level models with macro-level data,'' \emph{Journal of Building
  Performance Simulation}, vol.~6, no.~4, pp. 293--318, 2013.

\bibitem{YIN2016222}
R.~Yin, S.~Kiliccote, and M.~A. Piette, ``Linking measurements and models in
  commercial buildings: A case study for model calibration and demand response
  strategy evaluation,'' \emph{Energy and Buildings}, vol. 124, pp. 222 -- 235,
  2016.

\bibitem{EISENHOWER2012292}
{B. Eisenhower et al.}, ``A methodology for meta-model based optimization in
  building energy models,'' \emph{Energy and Buildings}, vol.~47, pp. 292 --
  301, 2012.

\bibitem{Chertkov2017}
\BIBentryALTinterwordspacing
M.~Chertkov, V.~Y. Chernyak, and D.~Deka, ``{Ensemble Control of Cycling Energy
  Loads: Markov Decision Approach},'' \emph{Working Paper}, pp. 1--21, 2017.
  [Online]. Available: \url{http://arxiv.org/abs/1701.04941}
\BIBentrySTDinterwordspacing

\bibitem{8442447}
M.~Chertkov, D.~Deka, and Y.~Dvorkin, ``Optimal ensemble control of loads in
  distribution grids with network constraints,'' in \emph{2018 Power Systems
  Computation Conference (PSCC)}, June 2018, pp. 1--7.

\bibitem{8515118}
{A. Hassan et al.}, ``Optimal load ensemble control in chance-constrained
  optimal power flow,'' \emph{IEEE Transactions on Smart Grid}, pp. 1--1, 2018.

\bibitem{Todorov2006}
E.~Todorov, ``{Linearly-solvable Markov decision problems},'' \emph{Advances in
  Neural Information Processing Systems}, vol.~19, no.~1, p.~8, 2006.

\bibitem{Reynders2014}
G.~Reynders, J.~Diriken, and D.~Saelens, ``Quality of grey-box models and
  identified parameters as function of the accuracy of input and observation
  signals,'' \emph{Energy and Buildings}, vol.~82, pp. 263--274, 2014.

\bibitem{DeConinck2014}
{De Coninck, R. et al.}, ``Rule-based demand-side management of domestic hot
  water production with heat pumps in zero energy neighbourhoods,''
  \emph{Journal of Building Performance Simulation}, vol.~7, no.~4, pp.
  271--288, 2014.

\bibitem{Patteeuw2014}
D.~Patteeuw and L.~Helsen, ``{Residential buildings with heat pumps, a verified
  bottom-up model for demand side management studies},'' in \emph{9th
  International Conference on System Simulation in Buildings}, no.~1,
  Li\`{e}ge, Belgium, December 10-12, 2014, pp. 1--19.

\bibitem{bruninx2018valuing}
{K. Bruninx et al.}, ``Valuing demand response controllability via chance
  constrained programming,'' \emph{IEEE Transactions on Sustainable Energy},
  vol.~9, no.~1, pp. 178--187, 2018.

\bibitem{Arteconi2015}
{A. Arteconi et al.}, ``Active demand response with electric heating systems:
  Impact of market penetration,'' \emph{Applied Energy}, vol. 177, pp. 636 --
  648, 2016.

\bibitem{Protopapadaki2014}
C.~Protopapadaki, G.~Reynders, and D.~Saelens, ``{Bottom-up modelling of the
  Belgian residential building stock: impact of building stock descriptions},''
  in \emph{9th International Conference on System Simulation in Buildings},
  vol.~2, no.~1, Li\`{e}ge, Belgium, December 10-12, 2014.

\bibitem{Baetens2015}
R.~Baetens and D.~Saelens, ``{Modelling uncertainty in district energy
  simulations by stochastic residential occupant behaviour},'' \emph{Journal of
  Building Performance Simulation}, vol. 1493, pp. 1--17, 2015.

\bibitem{Patteeuw2014JP}
{D. Patteeuw et al.}, ``{Integrated modeling of active demand response with
  electric heating systems coupled to thermal energy storage systems},''
  \emph{Applied Energy}, vol. 151, pp. 306--319, 2014.

\bibitem{cyx2011iee}
W.~Cyx, N.~Renders, M.~Van~Holm, and S.~Verbeke, ``{fIEE TABULA} typology
  approach for building stock energy assessment,'' VITO, Vlaamse instelling
  voor technologisch onderzoek, Tech. Rep., 2011.

\bibitem{EcofysStudyAppendix2013}
{K. Bettgenh\"{a}usernd et al.}, ``Heat pump implementation scenarios until
  2030, appendix,'' Ecofys, Tech. Rep., 2013.

\bibitem{Verhelst2012}
{C. Verhelst et al.}, ``Study of the optimal control problem formulation for
  modulating air-to-water heat pumps connected to a residential floor heating
  system,'' \emph{Energy and Buildings}, vol.~45, pp. 43--53, 2012.

\bibitem{Peeters2009}
{L. Peeters et al.}, ``Thermal comfort in residential buildings: Comfort values
  and scales for building energy simulation,'' \emph{Applied Energy}, vol.~86,
  no.~5, pp. 772--780, 2009.

\bibitem{Baetens2012}
{Baetens et al.}, ``{Assessing electrical bottlenecks at feeder level for
  residential net zero-energy buildings by integrated system simulation},''
  \emph{Applied Energy}, vol.~96, pp. 74--83, 2012.

\end{thebibliography}

\end{document}